\documentclass[a4paper, 12pt]{article}
\usepackage{epsfig}

\newcommand{\be}{\begin{equation}}
\newcommand{\ee}{\end{equation}}
\newcommand{\ba}{\begin{eqnarray}}
\newcommand{\ea}{\end{eqnarray}}
\newcommand{\bi}{\begin{itemize}}
\newcommand{\ei}{\end{itemize}}

\newcommand{\la}{\label}

\begin{document}
\begin{titlepage}
\begin{flushright}
\end{flushright}
\begin{centering}
\vfill

{\bf \large The Yang-Mills spectrum from a 2-level algorithm}

\vspace{1.5cm}

Harvey B. Meyer

\vspace{0.8cm}

Theoretical Physics, University of Oxford, 
1 Keble Road,\\ Oxford, OX1 3NP, United Kingdom\\
\vspace{0.2cm}
meyer@thphys.ox.ac.uk 
\vspace*{3.0cm}

\end{centering}

{\bf Abstract.-}
We investigate in detail a 2-level  algorithm 
 for the computation of 2-point functions of fuzzy Wilson loops
in lattice gauge theory. Its performance and
the optimization of its parameters  are described
in the context of 2+1D $SU(2)$ gluodynamics. 
In realistic calculations of glueball masses, it is found  
that the reduction in CPU time for given error bars on the correlator 
at time-separation $\sim0.2$fm, where a mass-plateau sets in, 
varies between 1.5 and 7
for the lightest glueballs in the non-trivial symmetry channels;
only for the lightest glueball is the 2-level algorithm not helpful.
For the heavier states, or for larger time-separations, 
the gain increases as expected exponentially in $m t$. 
We present further
physics applications in 2+1 and 3+1 dimensions and for different gauge groups
that confirm these conclusions.
\noindent 
\vfill

\vspace*{1cm}
\vfill
\end{titlepage}

\setcounter{footnote}{0}

\section{Introduction}

The study of pure gauge theories on the lattice at zero temperature has now
been for a few years in  an era of precision ``numerical experiments''. 
Highlights of the latter include
the determination of the low-lying glueball spectrum in 2+1~\cite{teper98}
and 3+1~\cite{mp} dimensions, the confining string spectrum (\cite{lw2} and~\cite{kuti}) as well as ratios of stable string tensions in $SU(N)$ gauge 
theories (\cite{debbio},~\cite{lucini}). The reasons for this progress lie
both in the increase in computing power and in the development of new 
numerical techniques, such as the fuzzing procedures 
(\cite{smear} and~\cite{block}), improved actions (e.g.~\cite{mp}) and 
multi-level algorithms (MLA) (\cite{multihit},~\cite{lw1} and~\cite{hbm}).
The purpose of this paper is to investigate the performance of the 
version proposed in~\cite{hbm}.

Retrospectively, we consider that 
the first multi-level algorithm was proposed  in~\cite{multihit}; 
it bears the name of the ``multihit method'', 
and consists in replacing the link
variables by their average under fixed nearest-neighbour links, when
computing a Wilson or Polyakov loop. It is a realisation of the
real-space renormalisation group transformation. 
Only much more recently
was it realised~\cite{lw1} that - thanks to the locality property - the idea
can be applied more generally and 
iteratively by performing nested averages under fixed 
boundary conditions (BC). 
Multi-level algorithms are of course particularly powerful in theories with 
a mass gap, where distant regions of the lattice are almost uncorrelated.
An impressive increase in the performance with respect to
 the ordinary 1-level algorithm was achieved in the Polyakov loop correlator
- the improvement is proportional to the area span by the two loops.
A further step towards generalization was taken in~\cite{hbm}, where the 
algorithm was adapted to any functional of the links - including fuzzy 
operators - that can be factorized. Indeed the factors need not even be
 gauge-invariant. The efficiency of the
algorithm is based on the fact that the UV fluctuations can be averaged out
separately for each factor, effectively achieving $n^{n_f}$ measurements
by only actually computing $n$ of them, 
where $n$ is the number of measurements done at the lower level of the 
algorithm and $n_f$ the number of factors. The choice of the factorization
is thus dictated by a competition between having as many factors as possible
and each factor being as independent of the BC as possible.

In glueball and string tension calculations, the smearing~\cite{smear} and 
blocking~\cite{block} techniques, used in conjunction with the variational
method~\cite{var}, have become part of the well-established machinery to
efficiently determine the glueball spectrum.
Such fuzzy operators have typically large overlaps onto the fundamental state
 in the studied symmetry channel and are far less sensitive to UV fluctuations
than bare Wilson loops. While it is clear that for asymptotically large 
Euclidean time separation, the multi-level algorithm 
becomes more efficient than the 1-level algorithm, 
the question of real practical interest is whether
one can truly improve the efficiency of realistic calculations. Typically,
the operators have reached mass plateaux already at 0.2fm in the case 
of glueballs. In such a regime, one cannot expect a statistical error 
reduction by orders of magnitude. Only a numerical  analysis can
reliably address the question formulated above.

It is equally important to determine whether the efficiency of the algorithm
 is maintained as the lattice spacing is decreased. Indeed the correlation
length becomes larger and larger in lattice units 
and one might wonder whether the low-level measurements at fixed BCs
 are still helping to reduce the dominant fluctuations on the correlator.

The outline of this paper is as follows.  After a description of the 
algorithm (section~\ref{description}), 
we investigate in section~\ref{parameters}
the dependence of the  2-level algorithm on its parameters 
in the context of glueball 2-point function calculations. 
We numerically demonstrate (section~\ref{performance})
that the parameters of the 
algorithm can easily be optimized  and compare its efficiency to that of the 
1-level algorithm as the continuum is approached. Special attention is paid
to the operators bearing vacuum quantum numbers, for which the vacuum 
expectation value (VEV) must be subtracted carefully. In 
section~\ref{physics}, we present
miscellaneous physics applications 
(glueballs and $k$-strings in 2+1 and 3+1 dimensions). 
We end with a summary and an outlook
on possible future developments. The appendix contains a 
comment on  the L\"uscher-Weisz algorithm.

\section{Algorithm description\la{description}}
In this section, we describe the implementation of a 2-level algorithm 
for the measurement of  2-point correlation functions. 
We use the isotropic Wilson action, and the 
update is done with compounds sweeps consisting of a
 1+3 mixture of heat-bath~\cite{kenpen} and 
over-relaxation~\cite{adler} sweeps.

The operators are smeared, blocked,
definite-momentum operators in 2+1  dimensional $SU(2)$ gauge theory. 
We shall use glueball operators as examples, 
however the conclusions will be shown in subsequent sections
 to be applicable to the measurement
of fuzzy spatial flux-tubes as well.

The details of the algorithm~\cite{hbm} are the following. 
The lattice size is $N_x\times N_y\times N_t$.
After a number $N_{up}$ of compound update sweeps, we freeze $N_t/\Delta$ 
time-slices separated by distance $\Delta$, and measure the average values
$\langle {\cal O}(t_i)\rangle_{bc}$
of the operators in all the other time-slices $t_i$ 
between the fixed time-slices by doing $n$ updates under these
fixed BCs. These average values 
in each time-slice are kept separately.
They are written to disk before updating the full lattice again.
$N_{up}$ is typically chosen to be $n/10$, so as to represent
a negligible amount of computer time, and nevertheless ensure good
statistical independence of the ``compound measurements'' (this will
be checked in section~\ref{performance}).

After $N_{bc}$ of these compound measurements, 
the correlator for $t\geq 2a$ can then easily be computed 
'off-line' from 
\be
\langle {\cal O}(t_i) {\cal O}(t_j)\rangle
= \frac{1}{N_{bc}}\sum_{bc} \langle {\cal O}(t_i)\rangle_{bc} 
\langle {\cal O}(t_j) \rangle_{bc},
\ee
if the time-slices $t_i$ and $t_j$ do not belong to the same ``time-block''.
This equation holds because 
the BCs have been generated with the weighting
given by the full lattice action~\cite{lw1}.

A few comments on the data storage are in order.
If $N_{op}$ are being measured, the amount of data generated is
\[ {\rm nb(data)_{II}} = N_{op} \times N_t \times N_{bc}. \]
This is to be contrasted with the ordinary way of storing the data:
the correlation matrix is computed during the simulation, and stored in 
typically $N_{bin}={\cal O}(100)$:
\[ {\rm nb(data)_{I}} \simeq N_{op}^2 \times N_{t} \times N_{bin}. \]
The ratio is thus
\be
\frac{\rm nb(data)_{II}}{\rm nb(data)_{I}} = \frac{1}{N_{op}}~~
 \frac{N_{bc}}{N_{bin}}
\ee
As an example, for a large production run with a total of $10^6$ measurements,
 we may do $n=10^3$ measurements under $N_{bc}=10^3$ fixed BCs.
Therefore, for $N_{op}\gg 10$ - which is usually the case -, the data size
is smaller than with the 1-level algorithm.
The obvious advantage of version II is that 
one can use a much larger number of operators (e.g. include non-zero momenta, 
scattering states, the square of the traces of operators,~\dots). 
Further advantages  include:
\begin{itemize}
\item one can  choose the binning \emph{a posteriori}, thus making a 
 more detailed check of auto-correlations possible;
\item if e.g. $\Delta=8$ and one is computing the 2-point function
at $t=5$, there are several ways to obtain it, which of course
all have the same average, but different variances; it is very convenient 
to be able to choose which combination is optimal \emph{a posteriori}
(see section~\ref{physics}).
\item 
in principle, one can extract 3- and 4-point function from the same data set,
as long as one correlates operators that have been averaged 
in different time-blocks.
Derivatives of the 2-point function can be computed just as easily.
\end{itemize}
On the other hand a downside of this way of proceding is that one looses
 information on the short-range correlator 
(0 and 1 lattice spacing of Euclidean time
separation). The time-zero correlator is useful because it allows one
to evaluate the overlap of the original operators onto the physical states.
In some cases it may be desirable to store the BC-averages of the short-range
 correlators  since 
\be
\langle {\cal O}(t_i) {\cal O}(t_j)\rangle = \frac{1}{N_{bc}}
\sum_{bc} \langle {\cal O}(t_i) {\cal O}(t_j)\rangle_{bc}
\ee
if the time-slices $t_i$ and $t_j$ belong to the same time-block. 
Incidently, for $t_i=t_j$, 
we shall see in section~\ref{performance}
that this measurement can be useful to predict  the 
performance of the algorithm for the larger time-separations.

\section{The algorithm and its parameters\la{parameters}}
We now present  data obtained at 
$\beta=12$, $V=32^3$ in the 2+1D $SU(2)$ pure gauge theory. 
Note that $\sqrt{\sigma}a = 0.1179(5)$~\cite{teper98}, 
which means that $a=0.055$fm 
(if we use $\sqrt{\sigma}=420$MeV) and we are indeed
well in the scaling region, as far as the low-energy observables are 
concerned.

We perform a check of the auto-correlation of compound measurements done 
at fixed BCs.
We then proceed to a study of the efficiency
of the algorithm as a function of its various parameters: first, the width
$\Delta$ of the time-block inside which the submeasurements are made;
secondly, the number of submeasurements. We will then look at 
the dependence of the error bars
on the mass of the state being measured.
\paragraph{Binning analysis}
On fig.~\ref{fig:binning}, we show the error bar on the correlator and its 
local-effective mass (LEM) 
for an operator lying in the $A_2$ irreducible representation 
 (IR, containing spins $0^-$, 4, 8, 12\dots) as function of the 
number of jacknife bins. We note that as long as the number of bins is not
much smaller than 100, the error bars are stable under the change of binnings.
Obviously the error bars are subject to fluctuations themselves, and in some
cases we will give estimates of the latter. However we can draw the lesson
than the number of updates $N_{up}\simeq n/10$ is apparently sufficient
to decorrelate the BCs sufficiently.
\paragraph{The distance between fixed boundary conditions}
Fig.~\ref{fig:a2_plain} 
show the dependence of the error bar on the $A_2$ correlator as a 
function of the number of submeasurements $n$, 
at rougly fixed CPU time. We consider that the comparison of error bars
 is meaningful at the $20\%$ level. The fundamental state
in that lattice IR is relatively heavy and known~\cite{regge} to have rotation
properties very similar to a continuum spin 4 glueball.
 The left graph shows the $\Delta=4$ data, the right one the $\Delta=8$ data.

In the first case, the
smallest error bar is achieved for the smallest number of measurements (here
$n=100$) for all time-separations ($ t=2,~3,~4$). On the 
right-hand side, the situation is the following: for time-separations 
$ t=2,~3,~4$, a small number of submeasurements ($n=200$) is
 more favourable, while, interestingly, the error bar for the 
 $ t =5$ correlator is practically independent of $n$.
For $ t \geq 6$, the hierarchy is inverted: the runs with a 
large number of subsweeps ($n=800,~1600$) yield smaller error bars.
This is consistent with the rule of thumb proposed in~\cite{hbm}, namely
that the optimal number of submeasurements should be of order $e^{mt}$.
In the present case, this evaluates to $\sim1300$, given that
 $am(A_2)\simeq1.2$.

We can already draw  the conclusion that $\Delta=4$ is too small a time 
block and a significant number of submeasurements does not
 lead to further variance reduction on the correlator. 
$\Delta=8 \simeq \frac{1}{\sqrt{\sigma} a}\simeq 
0.5{\rm fm}/a$ on the other hand 
seems well suited for that purpose. This conclusion is expected
to hold also in 3+1D pure gauge systems, since their long distance correlations
are very similar to the present 2+1D case.

It is also interesting to look at the error bars on the 
LEM. Indeed, when a large number of submeasurements is performed, the 
2-pt function at time $t$ and $t+a$ can be expected to be numerically 
more strongly correlated, thus leading to reduction of fluctuation of their 
ratio. This point is illustrated in fig.~\ref{fig:a2_plain}.
On the left (concerning $\Delta =4$), the  variance on the 
LEM at 3.5 lattice spacings is practically constant. On the right, we observe
that even at the smaller time separations $t = 2,~3,~4$, the 
runs with a large number of subsweeps (800) are at least as good as the 
runs with the smaller number of subsweeps (200). From 5.5 onwards, there
is a clear advantage at performing a large number of subsweeps. For instance, 
the error bar for the $n=800$ run is roughly 3 times smaller than that for 
the  $n=200$ run, and this at equal CPU time.
\paragraph{Time-separation dependence of the error bars}
It is also instructive to look at the same data from another point of view:
for a fixed number of submeasurements, how does the error bar on the 
correlator and on its LEM vary as a function of time separation?
On fig.~\ref{fig:a2_plain_YX}, it is clearly seen that the error bar 
decreases exponentially as the operators are measured further away from
the fixed boundaries. For $\Delta=8$, 
 the variance drops by a factor 100 between $t=2$ and $t=7$
for the runs with $n=800$ and 1600. 

\paragraph{Mass dependence of the error bars}
We plot the LEM as well as the local decay constant 
of the error bar on the correlator together on fig.~\ref{fig:LEM}.
 The upper figure illustrates the situation
with a large number of submeasurements ($n=1600$), 
while the lower shows what happens with only $n=200$.

We show two light states, the fundamental $A_1$ and $A_3$ states as well 
as the fundamental $A_2$ that was considered up to here.
 For the $A_1$ and $A_3$, the error bar decays along 
with the signal, since the former's decay constant matches the LEM 
of the corresponding operator.  As a consequence, long mass plateaux are
seen, with error bars increasing only very slowly. For the heavier $A_2$ 
state, the error bar decay constant keeps up only to 4.5 lattice spacings, 
resulting in a fast loss of the signal beyond that. It is 
nevertheless much more favourable a situation than with only 200 
submeasurements: while the lightest glueball plateau is obtained just as 
well, the $A_3$ data is much more shaky and the $A_2$ is essentially
 lost beyond 4.5 lattice spacings. We note that although the basis of $A_3$
 operators was the same for  $n=1600$ as for $n=200$, the variational
calculation performed slightly less well in the latter case.

In fact, the time separation where the error bar decay constant falls off 
on fig.~\ref{fig:LEM}
gives us an idea of the time-separation for which the number of 
submeasurements is optimal.
Indeed, if the error bar continues to fall off, it means that 
 the measurements have a large degree of statistical dependence
through the common BC,
since moving further away from the fixed BC makes them less dependent.
However, once -- far away from the BC --  the error bar is constant (i.e.
its decay constant is now zero), the signal to noise ratio is falling 
exponentially to zero.
Thus $n=1600$ is best suited for measuring the $A_2$ mass ($am\simeq 1.2$)
at 5.5 lattice spacings.

\section{Optimization procedure $\&$ performance\la{performance}}
We now proceed to a more systematic study of the efficiency of the 
2-level algorithm. We shall consider three states, in the $A_1$, 
 $A_2$ and $A_3$ lattice IRs. 
The lightest  states in these  
representations correspond~\cite{regge} to the 
$ J^P=0^+$, $J=4$ and $J=2$ continuum states.
The procedure we adopt is to measure these three correlators 
 at fixed physical Euclidean time separation $ t$. We do so at 
three values of $\beta=6,~9$ and 12 -- recall that in the scaling region, the 
lattice spacing simply scales as $1/\beta$. The correlator is evaluated for 
different numbers of submeasurements under fixed BCs:
\be
1\leq n \leq 200
\ee
We then plot the inverse efficiency $\xi^{-1}$ as a function of the number of 
submeasurements:
\be
\xi^{-1}(n) \equiv [\Delta C_n(t)]^2 ~\times~n, \la{effi}
\ee
where $\Delta C_n(t)$ is the error bar on the correlator when measured 
$n$ times under fixed BCs. In some cases, we shall also consider the 
efficiency with respect to the LEM, in which case $\Delta C_n(t)$ is replaced
by $\Delta m^{(\rm eff)}_n(t)$.

In this study the number of BCs was 100. They are separated
by 80  sweeps. The 
 individual measurements done under fixed BC were stored separately, to 
allow us to combine them in different ways. In particular, to obtain the 
efficiency corresponding to 10 submeasurements, for each BC
we can split the 200 submeasurements into
 20 'independent' sequences of 10 submeasurements. These 20 sequences are then
used to estimate the variance on the error bars themselves. 
On fig.~\ref{fig:a3},~\ref{fig:a2} and~\ref{fig:a1} we show 
these roughly estimated variances for $n\leq 20$, after what
the number of 'independent' sequences becomes smaller than 10 and these
variance estimates become unreliable. The aim here is only to give the order
of magnitude of the uncertainty on $\xi$, so as to be able to reach 
meaningful conclusions concerning its minimum as a function of $n$.

Eventually of course it is desirable to have an easier way to optimize the 
parameters of the algorithms. When we have an operator with exactly vanishing
vacuum expectation value (VEV),  we define a quantity $\omega$ as
the zero-time-separation correlator, multiplied by the number of 
submeasurements $n$:
\be
  \omega(n,t_i) = \frac{1}{N_{bc}}
\sum_{bc} ~~\sum_{{\rm meas}=1}^{n}
 \langle {\cal O}(t_i)^2 \rangle_{bc}
\ee
Obviously $\omega$ is a function of the distance between the time-slice
where the operator is measured and the fixed time-slices. 
It is easy to evaluate this quantity   accurately:
one of the objectives of this analysis is to check for the validity of this
 quantity as a predictor of the  optimal number of submeasurements
 of the 2-level algorithm. 
The absolute value of $\omega$ will not interest us, rather we will
check whether its minimum is reached at the same $n$ 
as $\xi^{-1}(n)$.
 
It  is also interesting to compare the efficiency of the 2-level algorithm to 
the standard 1-level algorithm with an equal number of  measurements.
In this case the translational invariance in the time direction is not 
broken by the algorithm. The sweeps between BCs have 
no \emph{raison d'\^etre} here; on the other hand 
 the measurements are done in each time-slice, 
including those that are kept fixed in the 2-level algorithm.
Thus the comparison of algorithms is fair.

Let us first consider the lightest $A_3$ state (fig.~\ref{fig:a3}).
The graphs correspond, from top to bottom to $\beta=6,~9$ and 12. We 
keep the physical time separation approximatively fixed at about 0.22fm
(2, 3, and 4 lattice spacings respectively), 
and similarly the separation of the fixed 
time-slices is augmented in lattice units ($4a$ at $\beta=6$, $6a$ at 
$\beta=9$ and $8a$ at $\beta=12$; 
we also show the case $\Delta=4a$ at $\beta=12$ for comparison). 
The first observation is that the 2-level algorithm
performs better at all three lattice spacings. If the number of 
submeasurements is chosen 'reasonably', the inverse efficiency is 
smaller by a factor $\sim3$ at the coarsest lattice spacing, and by a factor
$\sim2$ at both of the smaller lattice spacings, provided $\Delta$ is kept 
fixed in physical units. Secondly, the curve for $\xi^{-1}$ is extremely flat
around its minimum. For instance, at $\beta=9$ 
it seems that it does not matter whether one does 10 or 40 submeasurements, 
the performance for this particular observable will be unchanged. The flatness
 becomes even more pronounced closer to the continuum. This however is not
true for the case $\Delta=4a$ at $\beta=12$. Although the curve has a narrow
minimum at a small number of submeasurements, the efficiency then decreases 
rapidly and this setup becomes less favourable than the standard algorithm.
Thirdly, we note that the quantity $\omega$ shown at $\beta=12$ 
(it has been rescaled in such a 
way that it can be plotted along with the other curves) is a very good 
predictor of the minimum of the inverse efficiency curve $\xi^{-1}$, 
and this both
when $\Delta=4a$ and $\Delta=8a$. Its qualitative aspect (including the 
flatness) is very similar to the $\xi$ curve.

The qualitative statements that have been made for the $A_3$ correlator
 also apply to the $A_2$ correlator (see fig.~\ref{fig:a2}), whose mass
is larger by a factor $\sim4/3$. As one might expect, the higher mass 
favours the use of the 2-level algorithm even more: the gain in CPU time
for constant error bars is roughly a factor 6 at all three values of $\beta$.
Again the $\xi$ curve is extremely flat, but the optimal number of 
submeasurements has shifted to the right: in fact, 100 submeasurements
seems to be a good choice at all three lattice spacings.  Choosing a narrow
width for the time-blocks has the clear disadvantage of leading
to a smaller gain in efficiency and that this efficiency varies much more 
rapidly with the number of submeasurements. These facts are again 
well predicted  by the curve $\omega$.
\paragraph{The $0^{++}$ case}
We now move to the $A_1$ correlator, which gives the mass of the
 lightest glueball. Since this
is the trivial representation, the operator has a non-zero VEV, which has 
to be subtracted in one way or the other in order to extract information on 
the glueball spectrum. With the ordinary 1-level 
algorithm, it is customary to subtract the VEV \emph{a posteriori}:
\ba
C(t) & =& \sum_{t'}~\langle ({\cal O}(t')-\langle {\cal O}\rangle) 
({\cal O}(t+t')-\langle {\cal O}\rangle)  \rangle \nonumber\\
&\equiv& \sum_{t'=1}^{N_t}~\langle {\cal O}(t') {\cal O}(t+t') \rangle
- \frac{1}{N_t} \left(\sum_{t'=1}^{N_t} ~ \langle {\cal O}(t') \rangle\right)^2   
\label{vev}
\ea
This way of proceding is perfectly applicable to the 2-level algorithm, 
\emph{provided that only those measurements incorporated in
 the 2-point function are included in 
the VEV evaluation}. In other words, exactly the same measurements must 
appear in the second  sum as in the first in eqn.~\ref{vev}:
\be
C(t) = \sum_{t'\in\Theta_t}~\langle {\cal O}(t') {\cal O}(t+t') \rangle
- \frac{1}{\#(\Theta_t)} \left(\sum_{t'\in\Theta_t} ~ \langle {\cal O}(t') \rangle\right)^2 \la{vev2}
\ee
where $\Theta_t$ is a subset of  $\{1,\dots,N_t\}$. It varies with $t$:
depending the time-separation, the measurement of the correlator uses 
different time-slices. It is recommended to store the measurements
in  double precision, since the cancellation between the two sums grows 
with the time-separation.

Our experience is that failing to do the subtraction in this way leads
to a very large variance on the correlator (typically $30-50\%$ in a typical
run). The explanation is that in this way, one is really measuring, on a 
large but finite set of configurations, the fluctuation of the operator
around its average value \emph{measured on these configurations}.
Naturally, in the infinite statistics limit, both schemes give the 
same answer, but the proposed one benefits from the strong correlation
between the 2-point and 1-point function when they are measured on the same
 configurations.

There are of course many alternative possibilities\footnote{I thank Urs Wenger
for discussions on this point.}. One of them 
relies on the variational method~\cite{var}, which is widely used to improve
 the projection onto the fundamental state and to extract information on the 
excited spectrum. It was applied for instance in~\cite{regge} and consists 
in feeding  the \emph{unsubtracted} correlation matrices
 into the variational calculation. The generalized eigenvalue problem 
then yields the massless vacuum, followed by the fundamental glueball, the 
first excited, etc. The determination of the vacuum is very accurate in our 
experience, and the variance on the masses of the physical states did not
 seem to be higher. Naturally, one of the operators in the basis is wasted 
to project out the vacuum, but  this is not an issue when
one disposes of a large set of operators, as is usually the case.

Finally, we note that another group~\cite{majum} has used the 2-level 
algorithm for compact $U(1)$ 
scalar glueball calculations, where the forward-backward 
symmetric derivative of the correlator was taken. It is clear that at
small temporal lattice spacing, the finite-difference formula can evaluate
the derivative accurately, due to 
the large correlations between time-slices. The idea is thus related to 
that expressed by eqn.~\ref{vev2}.

These different methods are illustrated on fig.~\ref{fig:a1}:
 the inverse efficiency of the 1- and 2-level algorithms are plotted 
as a function of $n$. The VEV has been subtracted either by use of 
eqn.~\ref{vev2} or by applying the variational method to a set of three 
operators (the resulting operator had very large overlap onto the lightest
state in either method, and therefore a  comparison is meaningful).
We see that with either algorithm, the two VEV-subtraction methods perform
equally well. The second observation is that the 2-level algorithm is 
performing poorly here, if $n\geq10$. If we turn to the LEM,
we see that both the derivative-method and the direct VEV-subtraction method
have the same efficiency, once the number of measurements 
for $n\geq{\cal O}(50)$. For $n\leq50$, the VEV-subtraction looks better;
note however that, for discretization reasons,
 the LEM on the derivative is actually at 4 lattice spacings, rather than 3.5.

\section{Physics applications\la{physics}}
We present four datasets: glueballs  and $k$-string
tensions in 2+1 and 3+1 dimensions. We shall be comparing the efficiency
of the ordinary 1-level algorithm with the  2-level algorithm.
A general comment applies to all four cases: the operators used are all
variationnally~\cite{var} determined linear combinations of
smeared~\cite{smear}, blocked~\cite{block} operators which large overlaps 
onto the physical states. The variational method involves 
Cholesky-decomposing a correlation matrix, which fails if the statistical 
noise spoils the positivity of the matrix. Our experience with the 1-level
algorithm is that the procedure often fails for that reason if it is applied 
at $t\geq1$. By contrast, we were always able to perform the decomposition 
when using the 2-level algorithm. The explanation is that the operators 
decaying more rapidly, which couple to  excited states, 
are more accurately measured in such a way that the positivity is preserved.
\subsection{Glueballs in 2+1 dimensions}
As a first physics application of the methods that have been presented,
we  extract the masses of several excited-glueball masses in 2+1D $SU(2)$ 
gluodynamics. Table~\ref{g2+1} shows the data at three different lattice 
spacings. The parameters of the 2-level algorithm are $n=5000,~1000,~800$
and $\Delta=4,~6,~8$ respectively for $\beta=6,~9,~12$.
The variational method is applied at 2.5 lattice spacings for the 
first two lattice spacings, and 3.5 lattice spacings for the third.
A detailed  efficiency comparison has  already been made in 
section~\ref{performance}. The identification of the continuum numbers
was worked out in~\cite{regge} and~\cite{hspin}, and will be further
detailed elsewhere~\cite{dphil}, but we now dispose of increased statistics
and make  use of the 2-level algorithm: this allows us to 
obtain an accurate determination of glueball masses as heavy as three times
the lightest glueball. In the future this ability will have to 
be followed by
the  development of  techniques to reduce systematic uncertainties on 
glueball spectroscopy measurements (mixing of single-glueball states 
with scattering states, multi-torelon states,~\dots).

\subsection{Glueballs in 3+1 dimensions}
The multi-level algorithm applied to gauge-invariant 
correlators was first tested on glueball operators in the
3+1 dimensional $SU(3)$ gauge theory~\cite{hbm}. Some exploratory steps  were
carried out to optimize the parameters, while in this paper
 we have performed a much more detailed study in 2+1 dimensions. 
One might wonder whether the conclusions carry over to 3+1 dimensions, since
the short-distance fluctuations scale differently. 

Here we shall simply present a comparison of efficiency
in a  realistic case of glueball calculations at $\beta=6.0$,  $\beta=6.2$
and $\beta=6.4$, where we can compare our data to that of the 10-year-old
UKQCD data~\cite{ukqcd}. 
The parameters of the 2-level algorithm are $n=40$ for all three values of 
$\beta$, while $\Delta=8$ for $\beta=6.2$ and 6.4, and $\Delta=6$ for
$\beta=6.0$.
Let us focus on the lightest states in the 
$A_1^{++}$, $E^{++}$ and $T_1^{++}$ representations (see table~\ref{g3+1}). 
We compare the efficiency in terms of the error bars on the
LEMs by scaling the 1-level error bar to the number of 
sweeps done in the run where the 2-level algorithm was implemented (see
eqn.~\ref{effi}).
The same conclusions hold than in 2+1D: apart from the lightest glueball, 
the efficiency of the 2-level algorithm is greater than that of the 1-level
one, and increases rapidly with the mass of the state.
The comparison to the UKQCD data is slightly less robust, because the 
operators used are not the same. The difference in the extent of the time
direction was compensated by scaling up the statistics of the 2-level run.
Nevertheless, the same trend is observed as in the comparisons at the coarser
lattice spacings.

Consider the correlator at four lattice spacings. It can be obtained 
by correlating the time slices situated symmetrically around the fixed 
time-slice, or asymmetrically. Naturally, the first way is more favourable.
However, for a very massive state, the measurements are expected to
be very weakly correlated to the fixed BC; therefore the asymmetric
correlator can increase the statistics and reduce the final error bar.
In fact, one can make any mixture of both measurements. If $\bar t$ is the 
time-coordinate of the fixed BC:
\ba
C(t=4)& \propto& \frac{\alpha}{N_{bc}}\sum_{bc}
   \left[{\cal O}(\bar t+1){\cal O}(\bar t-3)
+ {\cal O}(\bar t-1){\cal O}(\bar t+3)\right]\nonumber  \\
&+& \frac{1}{N_{bc}}\sum_{bc}{\cal O}(\bar t+2){\cal O}(\bar t-2)
\ea
The parameter $\alpha$ can be optimized as well. This is illustrated 
on fig.~\ref{fig:a1var}, where the variances of the LEMs are plotted as 
a function of $\alpha$ for $\beta=6.0$. We see that the optimal value of 
$\alpha$ increases with the mass of the state, as one might expect. However, 
the dependence on $\alpha$ is weak for $\alpha\geq0.2$: choosing $\alpha=0.3$
is practically optimal for all states and $\alpha=1$ is not much worse
(note that the choice of $\alpha$ is done \emph{a posteriori}).

\subsection{$k$-strings in 2+1 dimensions}
We present an excerpt of a  dataset that will be published in full 
elsewhere~\cite{altes}. We compute masses of $k=1,~2,~3$ and 4 fuzzy 
flux tubes in 2+1D $SU(8)$ gauge theory; 
our methodology follows that of~\cite{lucini}. Table~\ref{k2+1} contains data
 obtained both with the 1- and 2-level algorithms on a relatively 
coarse lattice ($\sqrt{\sigma}a=0.2550(4)$). 
The algorithm parameters are $n=1000$, $\Delta=4$.
It was anisotropic in size in order to check for corrections to the linear
 dependence of the mass on the length of the Polyakov loop.

The small mass of the $k=1$, $L\simeq2$fm 
state makes the 1-level far more efficient
 to compute the LEM even at 2.5 lattice spacings. The 2-level algorithm
only becomes superior for $L\geq 2.6$fm. 
Another point is that the overlap of our operator onto the fundamental 
state is better than $99\%$, which means that the mass can be extracted
 already at 0.5 lattice spacings. 
The lesson we learn from this is that it is worth keeping the 
correlator at small time-separations, as indicated in 
section~\ref{description}.

For all other states, the conclusions are quite different: the 2-level
algorithm is 4 times more efficient for the $k=2$ $L=2$fm string, and 
this gain in efficiency grows extremely rapidly with the mass of the
state.

\subsection{$k$-strings in 3+1 dimensions}
Finally, we present  data obtained some time ago on $k=1$ and $k=2$ strings 
in 3+1D $SU(4)$ gauge theory at three different lattice spacings 
(see table~\ref{k3+1}). For $\beta=10.90,~11.10,~11.50$,
 the parameters of the 2-level algorithm were $n=80,~20,~20$
and  $\Delta=6,~4,~4$ respectively. The masses are extracted from cosh fits
starting at 2.5 lattice spacings. The data at the first two values of $\beta$
can be directly compared to that of~\cite{lucini}, 
where $10^5$ sweeps were done in each case with the 1-level algorithm. 
Two comments are in order: the $k=1$ string is more accurately obtained
with the 1-level algorithm, while the 2-level algorithm performs slightly
better for the $k=2$ string at $\beta=11.10$, 
although its parameters may not have been optimal. At $\beta=10.90$, we notice
a discrepancy between the two masses of  2.9 standard deviations. 
This is presumably due to the fact that the
cosh fit was started at one lattice spacing~\cite{mike} for that particular
state in the work of Teper and Lucini. Generally speaking,
the 2-level algorithm has the advantage of yielding longer mass plateaux
because of the decrease of the error bar with the separation to the BC, 
and therefore helps reducing the systematic error. 
Indeed lattice calculations, because of the positivity 
property of correlators, naturally
tend to overestimate the masses extracted from them.

\section{Summary $\&$ outlook}
It is time to summarize what we have learnt about  the 2-level algorithm.
We have emphasized the linear dependence of the data size on the number of 
operators; auto-correlations can be checked for easily, and the precise
way in which the correlator is computed can be optimized \emph{a posteriori}.

The optimization study of the parameters led to the conclusion that
$\Delta\simeq\frac{1}{\sqrt{\sigma}a}$ is a good choice for the separation
of the fixed time-slices. In that case,  the variance of the correlator
decreases exponentially $\sim e^{-mt}$ as long as the number of measurements
at fixed boundary conditions $n>e^{mt}$. As a consequence, longer mass 
plateaux are seen, even for the more massive states. This feature should
help in reducing the systematic bias to overestimate the masses 
being calculated.

Suppose we want to compute the correlator at time-separation $t$
from measurements in time-slices $\bar t+t/2$ and $\bar t-t/2$ with respect
to the fixed time-slice position $\bar t$.
The optimization of $n$ can be achieved by minimizing 
  [the $t=0$ correlator measured $n$ times at distance $t/2$
from fixed time-slices] $\times~n$. This  is an easy quantity to compute
as function of $n$; it is sufficient to  store
the individual measurements  separately. For a fixed physical separation 
$t$, the optimal number of measurements $n$ is only weakly
dependent on the lattice spacing. A possibility that we have not explored
is to let the number of measurements depend on the boundary conditions, 
with a termination condition determined by the  desired accuracy (which
would presumably be chosen to be proportional to $1/\sqrt{N_{bc}}$).

The efficiency of the 2-level algorithm was compared to that of the 1-level
algorithm in 2+1 and 3+1 dimensions for various gauge groups. 
We found that the 2-level algorithm performs better
 for all glueball states but the lightest. The kind of gain in computing-time
 one can expect in realistic glueball spectrum calculations
varies between 1.5 and 7 for the lightest states
in the lattice irreducible representations in the case of  2+1D $SU(2)$. 
The gain then increases exponentially with the mass of the state.
If high accuracy is required for the lightest glueball, it might make 
sense to do a separate run using the 1-level algorithm: at any rate, it will
use far less computing time than is required for the heavy states.
The same qualitative statements apply in computations of flux-tube masses.
The 2-level algorithm starts to become more favourable at a string length
of $\sim2.5$fm; and it is always more performant for the excited states and 
the  strings of higher representations.

We would like to conclude by mentioning two further applications 
of the 2-level algorithm. As was suggested in~\cite{hbm}, 
the method should be well suited to compute 3-point functions
of glueballs~\cite{michaelg} and flux-tubes~\cite{michaelft}, 
since these observables involve 3 factors, each subject to UV fluctuations. 
The possibility is being investigated.

An alternative to variational calculations
in conjunction with a large number of fuzzy operators is the  spectral
function method (for a review, see~\cite{spec}) in conjunction with the 
maximal entropy method to perform the inverse Laplace transform.
It would be  interesting to investigate the possibility of 
using  UV operators (e.g. a bare plaquette, which couples equally 
to many states) to extract the glueball spectrum. The
correlator would need to be measured very accurately -- and here we
expect  the 2-level algorithm to be of great help --
on a lattice with a very fine temporal resolution.
We leave this line of research open for the future.

\paragraph{Acknowledgements} In the course of this work I 
benefited a lot from interacting with 
 Biagio Lucini, Michael Teper and Urs Wenger  at Oxford University.
I also thank M.~Hasenbusch, S.~Kratochvila, P.~Majumdar and P.~Weisz for 
interesting discussions at LATTICE 03. 
The numerical calculations were performed on PPARC and EPSRC
funded workstations and on a Beowulf cluster in Oxford Theoretical Physics.
Finally I wish to thank  the Berrow Trust for financial support.


\vspace*{1cm}

{\Large \bf Comment on the L\"uscher-Weisz algorithm}
\vspace{0.3cm}\\
Consider the LW algorithm for bare time-like Polyakov loop correlators.
It was noted by several groups 
(\cite{weisz},~\cite{hasen},~\cite{dvshif} (where the implementation is due 
to K.~Rummukainen)) that the LW algorithm  can probably still be 
improved (and also simplified) for large enough separation $R$.

Indeed the physical system consisting of one 
``time block'' with fixed spatial links as BC  can be studied for
itself. It is reminiscent of the finite-temperature gauge system, where the 
BC are however periodic. The two systems have the same exact 
$Z_N$ symmetry which can be broken spontaneously if the width of the
 block is small (corresponding to high temperature). It must thus be ensured
that this does not happen, otherwise the error reduction will fail to take 
place: the VEV taken by the segments of Polyakov loop will only be averaged
out at the outer level, and will therefore suffer 
from ${\cal O}(1/\sqrt{N_{bc}})$ fluctuations.

The BCs break Lorentz invariance in the subsystem. However, by analogy with
Dirichlet BCs, we expect that the dimensionally-reduced
theory, containing a adjoint Higgs field and the gauge field, exhibits a 
pseudo-mass gap. In this situation, the correlation between the two segments
of link products decays very rapidly at large distance $R$ 
 and it should help to perform multiple
 measurements of the two segments separately by keeping a slice $S$
 between them fixed. In this way it is not necessary to store the direct
 product of $SU(N)$ matrices. In fact, this version of MLA has been found 
to perform well even 
in the presence of an adjoint field (\cite{dvshif}).

\newpage
\begin{table}
\begin{center}
\begin{tabular}{|c|c|c|c|c|c|}
\hline
IR    & spin & $\beta=6.0$ & $\beta=9.0$ & $\beta=12.0$ & continuum: \\
   &   & \tiny{$a\sqrt{\sigma}=0.2538(10)$} & 
\tiny{$a\sqrt{\sigma}=0.1616(6)$}  
 & \tiny{$a\sqrt{\sigma}=0.1179(5)$}   & $m/\sqrt{\sigma}$\\
& & $10^7$ sweeps &$3\cdot10^6$  sweeps&$3\cdot10^6$  sweeps& \\
\hline 
$A_2$ & 4    & 2.423(34)   & 1.5766(96)  & 1.183(16)   &  10.01(16) \\
      & 4    & 2.638(85)   & 1.804(42)   & 1.395(57)   &  11.91(45) \\
      & 4    & 2.89(11)    & 1.856(86)   & 1.468(77)   &  12.22(69) \\
      & 4    & 3.01(15)    & 2.199(44)   & 1.698(44)   &  14.98(49) \\
\hline 
$E$   & 3    & 2.552(70)   & 1.663(23)   & 1.2587(51)  & 10.84(14) \\
      & 1    & 2.558(43)   & 1.813(20)   & 1.360(13)   & 11.95(18) \\
      & 3    & 2.718(63)   & 1.855(28)   & 1.350(23)   & 11.78(27) \\
\hline 
\end{tabular}
\end{center}
\label{g2+1}
\caption{The lightest states in the $A_2$ and $E$ lattice irreducible
representations. The string tension values are taken from~\cite{teper98}.
For 2+1D $SU(2)$ on $16^3,~24^3$ and $32^3$ lattices respectively for the
three values of $\beta$.}
\end{table}

\begin{table}
\begin{center}
\begin{tabular}{|c|c|c|c|}
\hline
$\beta=6.0$& $m^{\rm 1-lev}_{\rm eff}(2.5a)$ & $m^{\rm 2-lev}_{\rm eff}(2.5a)$ &
\large{$\frac{\xi_{\rm 2-level}}{\xi_{\rm 1-level}}$ }\\
 $16^3\times36$   & $4.16\cdot10^5$ sweeps&  $15.04\cdot10^5$ sweeps & \\
\hline
$A_1^{++}$& 0.7106(87) & 0.7248(55) &  0.69\\
$E^{++}$  & 1.078(16)   & 1.0776(63) & 1.80 \\
$T_1^{++}$&  1.605(90)  &  1.612(18) & 6.55 \\
\hline
\end{tabular}
\vspace{0.6cm}\\
\begin{tabular}{|c||c|c||c|}
\hline
$\beta=6.2$& $m^{\rm 1-lev}_{\rm eff}(3.5a)$ & $m^{\rm 2-lev}_{\rm eff}(3.5a)$ &
\large{$\frac{\xi_{\rm 2-level}}{\xi_{\rm 1-level}}$} \\
 $24^3\times32$       & $2\cdot10^5$ sweeps&  $9.28\cdot10^5$ sweeps & \\
\hline
$A_1^{++}$ & 0.531(12) & 0.5273(62) &  0.83 \\
$E^{++}$   & 0.768(22)  & 0.7819(64) &  2.46 \\
$T_1^{++}$  & 0.99(15) & 1.250(27)  &  6.60\\
\hline
\end{tabular}
\vspace{0.6cm}
\begin{tabular}{|c||c|c||c|}
\hline
$\beta=6.2$& $m^{\rm 1-lev}_{\rm eff}(2.5a)$ & $m^{\rm 2-lev}_{\rm eff}(2.5a)$ &
\large{$\frac{\xi_{\rm 2-level}}{\xi_{\rm 1-level}}$} \\
  $24^3\times32$       & $2\cdot10^5$ sweeps&  $9.28\cdot10^5$ sweeps & \\
\hline
$A_1^{++}$ & 0.5269(77) &  0.5369(52) &  0.48 \\
$E^{++}$  & 0.8079(99)   & 0.8026(43) & 1.12  \\
$T_1^{++}$  & 1.260(39) & 1.294(11)  & 2.31 \\
\hline
\end{tabular}
\vspace{0.6cm}
\begin{tabular}{|c|c|c|c|}
\hline
$\beta=6.4$ & $m^{\rm UKQCD}_{\rm eff}(2.5a)$~\cite{ukqcd}
&$m^{\rm 2-lev}_{\rm eff}(2.5a)$ &   
\large{$\frac{\xi_{\rm 2-level}}{\xi_{\rm 1-level}}$}   \\
    & $V=32^4$:  $0.322\cdot10^5$ sw 
&$V=32^3\times48$: $1.11\cdot10^5$ sw &  \\
\hline
$A_1^{++}$&0.415(14) &  0.4000(73) &  0.64 \\
$E^{++}$  & 0.620(17)&   0.5894(72)& 1.08\\
$T_1^{++}$& 1.06(8)  &   0.946(10) & 12.4\\
\hline
\end{tabular}
\end{center}

\caption{Comparison of local effective masses 
 using the ordinary 1-level and the 2-level algorithms in 3+1D $SU(3)$. 
The ratios of efficiencies $\xi$,
representing the inverse ratio of CPU time 
required for fixed accuracy,  is given in the last column. In the last
case, the statistics of the 2-level run were scaled up by 1.5 in the 
efficiency computation to take the different volume into account.\label{g3+1}}
\end{table}

\begin{table}
\begin{center}
\begin{tabular}{|c|c|c@{}c|c|c|}
\hline
string& state & $m_{\rm eff}(1.5a)$ &
$m_{\rm eff}(2.5a)$
&$m^{\rm 2-level}_{\rm eff}(2.5a)$    & 
\large{$\frac{\xi_{\rm 2-level}}{\xi_{\rm 1-level}}$ } \\
length  &    &1-level:   &   $1.2\cdot10^5$ sweeps   
&$16\cdot10^5$ sweeps &  \\
\hline
& $k=1$& 1.0130(82)& 1.001(21)& 1.003(16) &  0.13\\
$L=16$& $k=2$& 1.739(37) & 1.74(22) & 1.729(30) & 4.0\\
& $k=3$& 2.24(10)  & /        & 2.216(66) & / \\
& $k=4$& 2.63(25)  &  /       & 2.32(12)  &  /\\
\hline
&$k=1$&   1.292(17) & 1.359(66)   & 1.288(19) &  0.90\\
$L=20$ &$k=2$& 2.27(11)  &  1.92(77)   &  2.174(59)&  13 \\
&$k=3$& 2.63(27)& /   &  2.53(23)&   /\\
&$k=4$& 3.01(72)  & /   & 3.04(33) &  /  \\
\hline
\end{tabular}
\end{center}
\caption{The 4 $k$-string tensions in 2+1D $SU(8)$ gauge theory, 
$V=16\times20\times24$, $\beta=115.0$, measured with two different
 algorithms ($n=1000$,~$\Delta=4$ for the 2-level case).}
\la{k2+1}
\end{table}

\begin{table}
\begin{center}
\begin{tabular}{|c|c|c|c|}
\hline
state &   $\beta=10.90$ & $\beta=11.10$ &  $\beta=11.50$  \\
       &$ L=12 $  & $ L=16 $  &      $ L=24 $ \\
   &$1.7\cdot10^5$ sweeps  & $0.83\cdot10^5$ sweeps & $3.6\cdot10^5$  sweeps\\
\hline
$am_1$      &  0.610(19)    &    0.592(13)     & 0.4638(71)\\  
$am_1^{*}$  &   0.728(75)    &   0.700(25)      &  0.680(20)\\
\hline
$am_2$ &      0.823(25)    &  0.820(17)       & 0.616(16) \\%
$am_2^{*}$&   1.187(81)    &  0.991(56)       & 0.913(35) \\%
\hline  
$\sigma_1 a^2$ & 0.0581(18)& 0.04109(90) &  0.02114(32) \\
$\sigma_2 a^2$ &  0.0759(23)  & 0.0553(11) & 0.02748(71)  \\
$\sigma_2/\sigma_1$&  1.306(56) & 1.346(40)     & 1.300(39)     \\
\hline
\end{tabular}
\end{center}
\caption{The masses of $k=1$ and $k=2$ strings for 3+1D $SU(4)$ gauge theory;
the string tensions are computed assuming the L\"uscher correction
$m=\sigma L-\frac{\pi}{3L}$, and in the error bar on the ratio
 $\sigma_2/\sigma_1$ statistical independence is assumed.}
\la{k3+1}
\end{table}
\begin{figure}[tb]

\centerline{\begin{minipage}[c]{14cm}
    \psfig{file=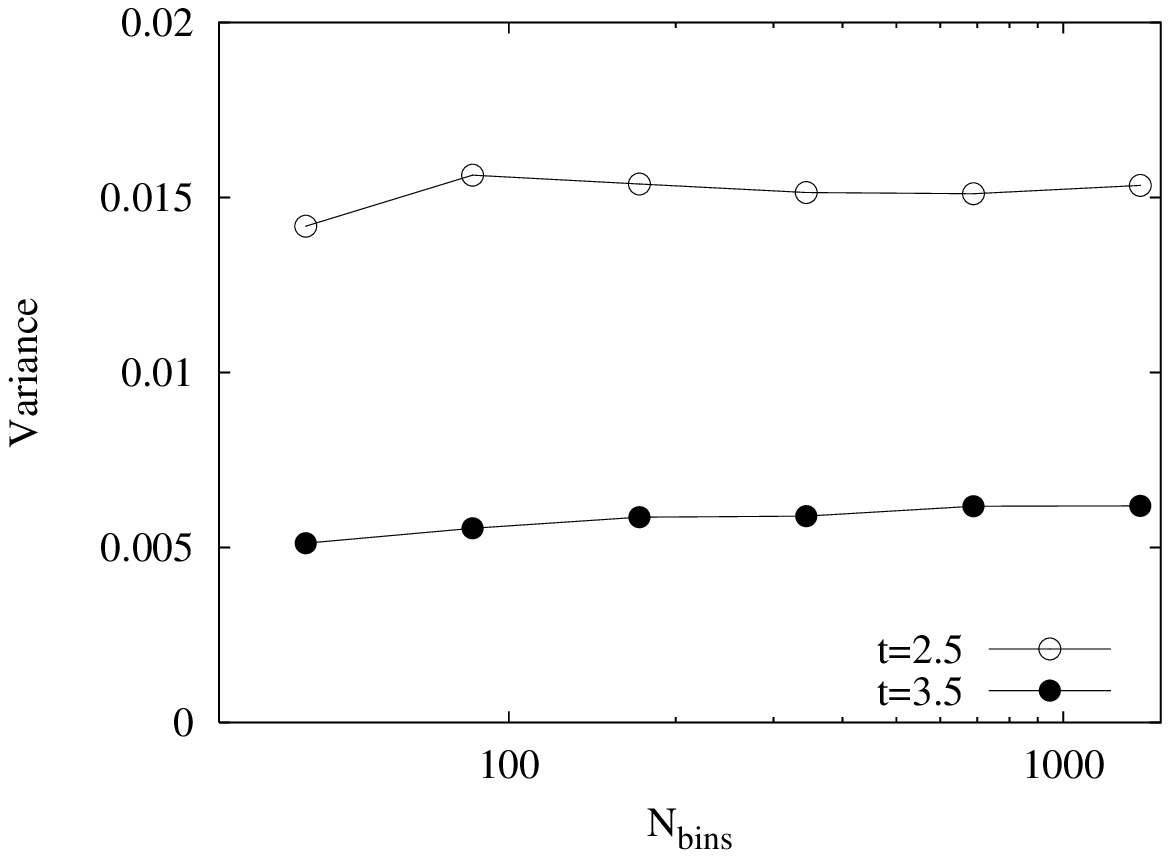,angle=0, height=9cm}\\
	\psfig{file=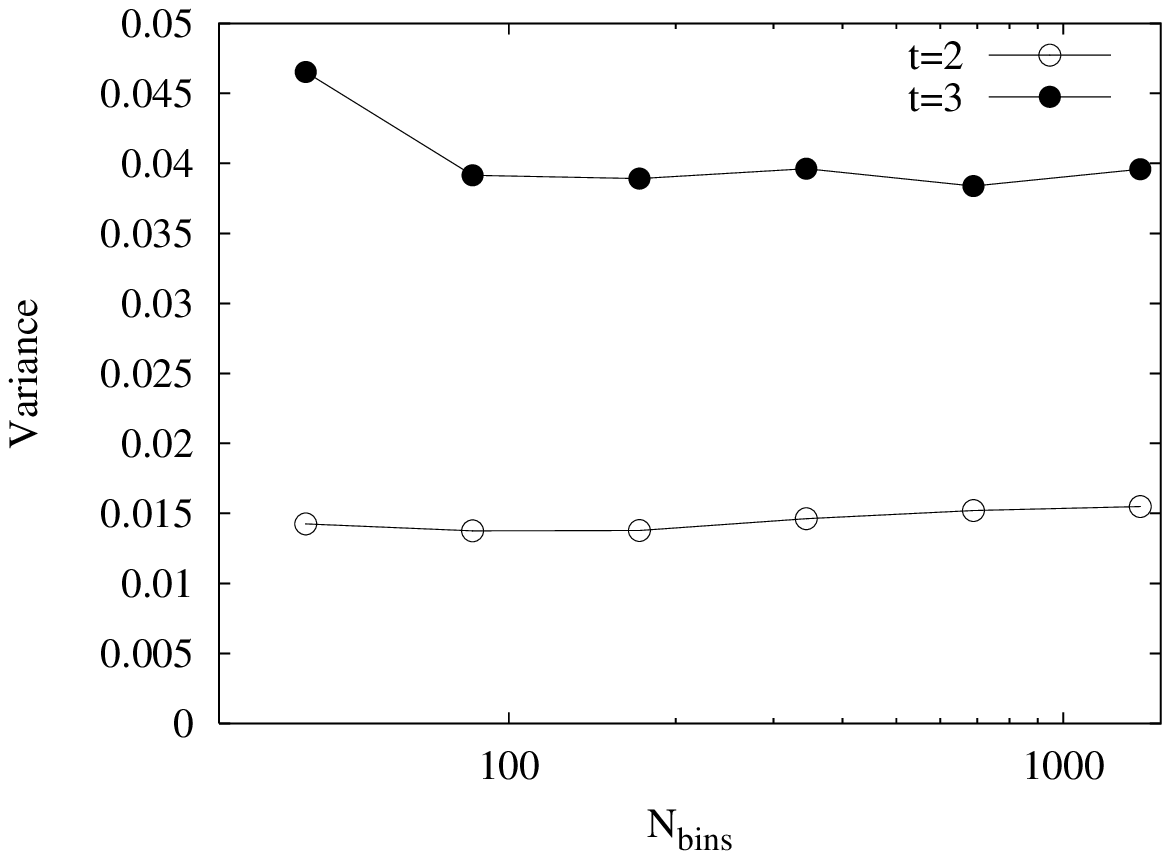,angle=0, height=9cm}	
    \end{minipage}}
\vspace*{0.5cm}

\caption[a]{Jacknife-bin-size dependence of the statistical error on 
the $A_2$ correlator (top) and its local-effective-mass (bottom). The 
separation of the fixed time-slices is  $ \Delta = 4$. For 2+1D $SU(2)$, 
at $\beta=12,~V=32^3$ and $n=100,~N_{bc}=1400$.}

\la{fig:binning}
\end{figure}

\begin{figure}[tb]

\centerline{\begin{minipage}[c]{16cm}
    \psfig{file=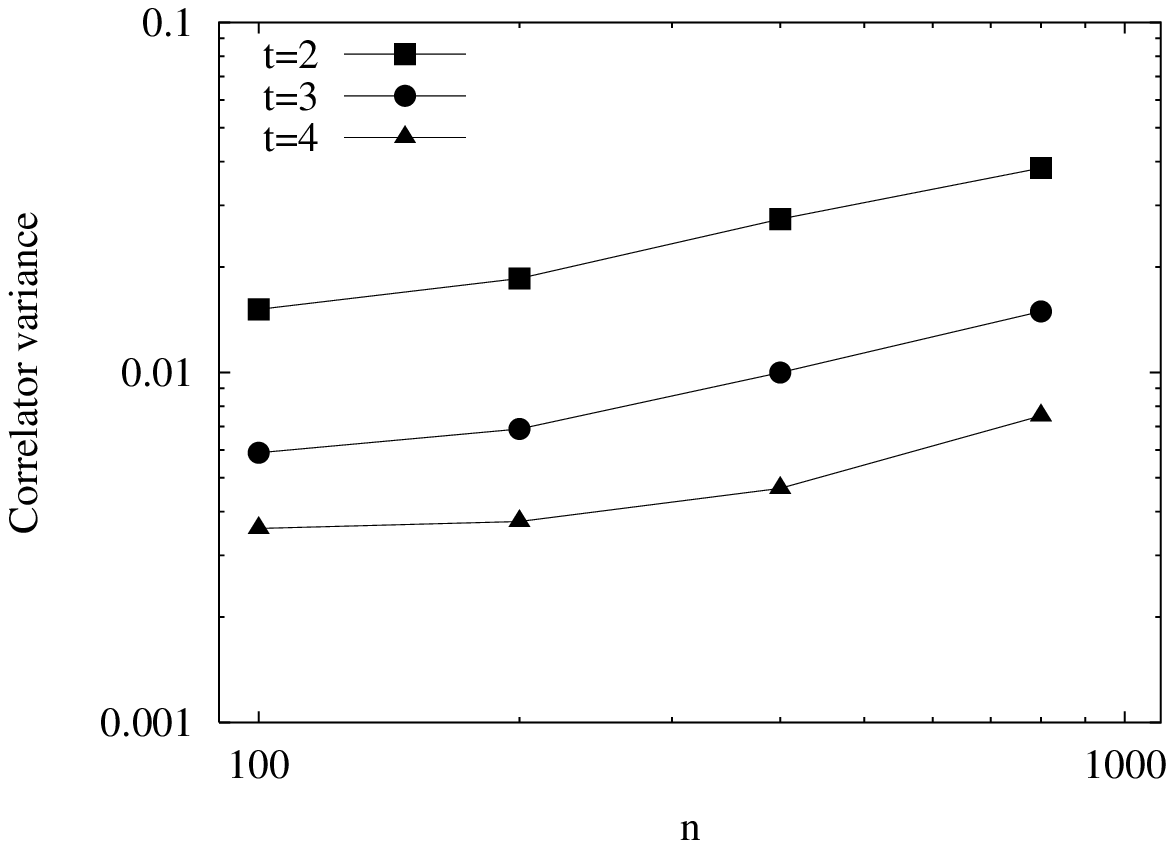,angle=0,width=8cm}
	\psfig{file=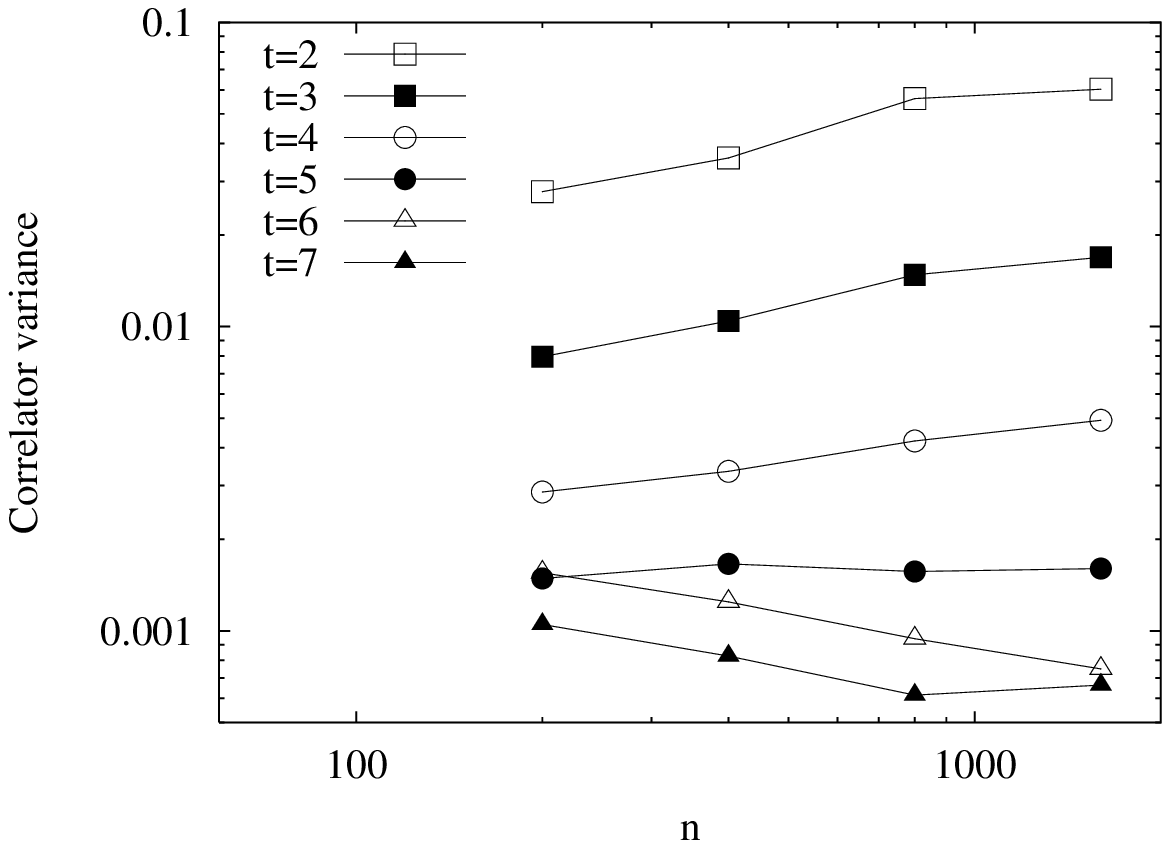,angle=0, width=8cm}
      \psfig{file=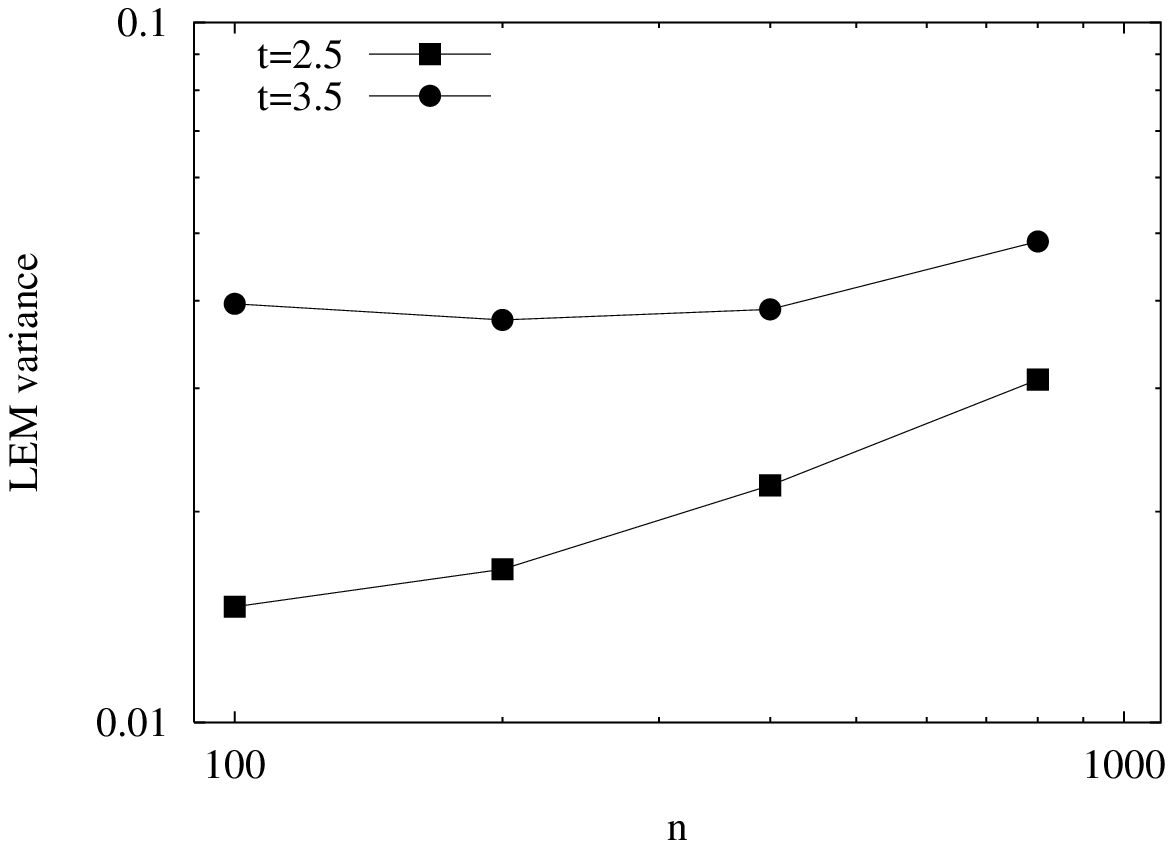,angle=0,width=8cm}
	\psfig{file=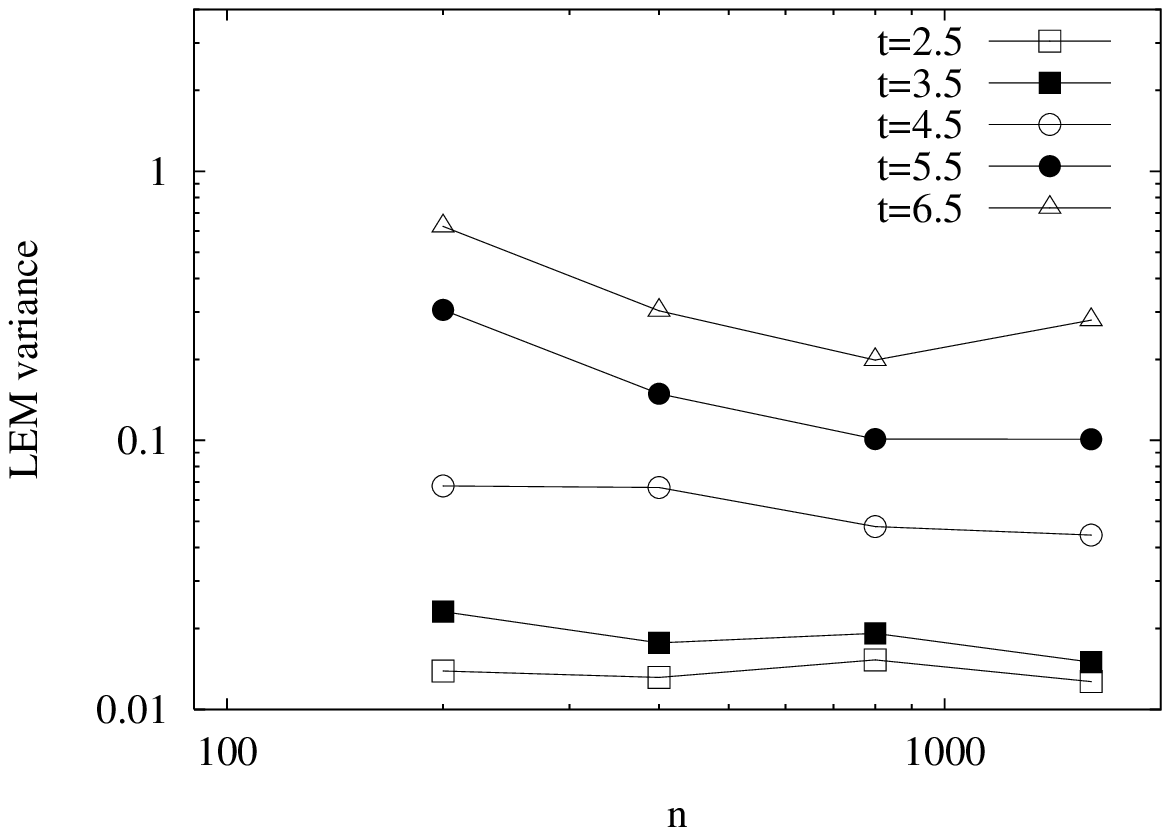,angle=0,width=8cm}
    \end{minipage}}
\vspace*{0.5cm}

\caption[a]{The variance of the correlator (top) and the local effective mass
 (bottom), as function of  the number of 
measurements under fixed boundary conditions $n$, for fixed computing
time. The separation of the fixed time-slices is $\Delta  =4 $ on the left
 and  $\Delta  = 8 $ on the right. The operator is a linear combination of
fuzzy magnetic Wilson loops lying in the $A_2$ square lattice irreducible
representation. For 2+1D $SU(2)$, at $\beta=12,~V=32^3$.}

\la{fig:a2_plain}
\end{figure}

\begin{figure}[tb]

\centerline{\begin{minipage}[c]{16cm}
\psfig{file=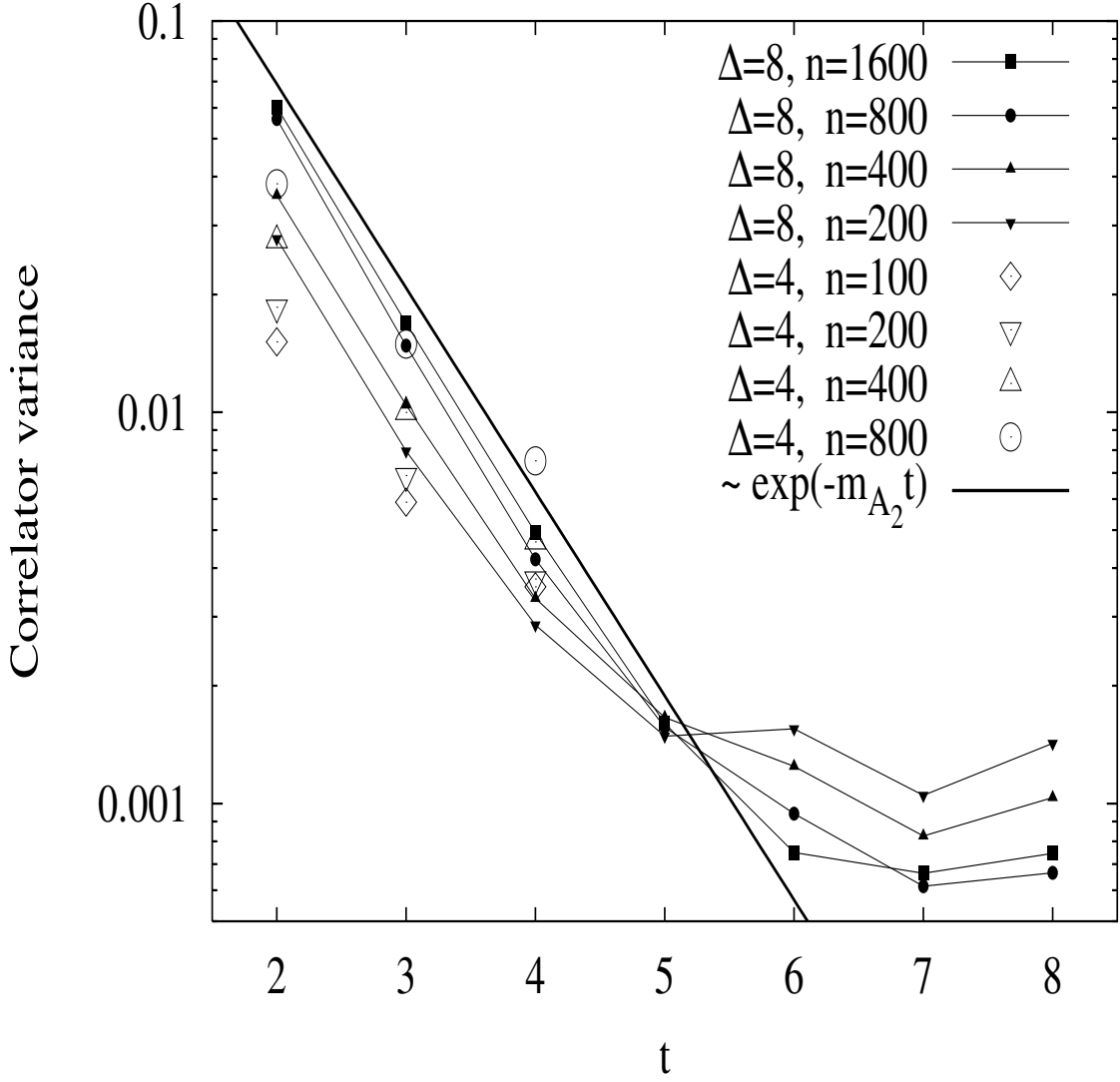,angle=0,height=15cm, width=16cm}
    \end{minipage}}
\vspace*{0.5cm}

\caption[a]{$A_2$-correlator variance as function of  the Euclidean-time
 separation $t$, for  different numbers of measurements under fixed
boundary conditions $n$ and separation of the fixed time-slices $\Delta$. 
The computing time is the same for all points.
For 2+1D $SU(2)$, at $\beta=12,~V=32^3$.}
\la{fig:a2_plain_YX}
\end{figure}

\begin{figure}[tb]

\centerline{\begin{minipage}[c]{14cm}
    \psfig{file=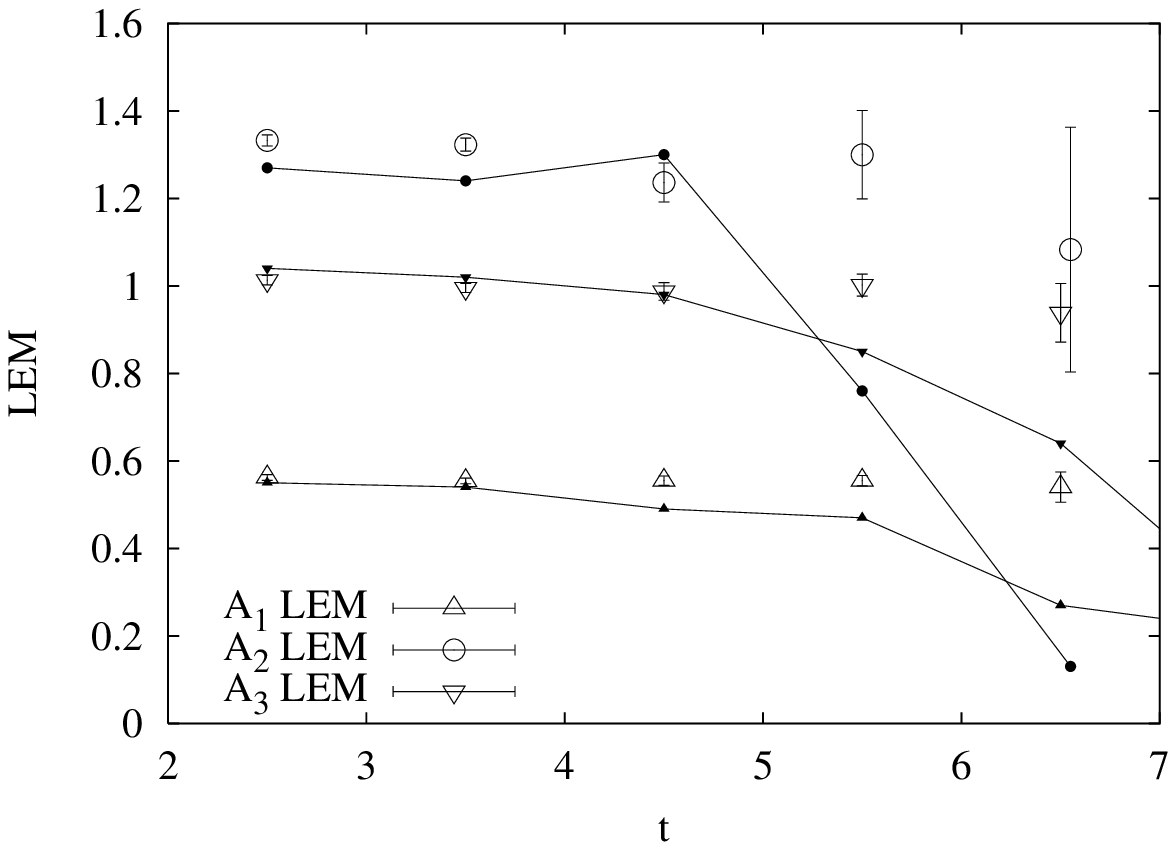,angle=0, height=9cm}\\
	\psfig{file=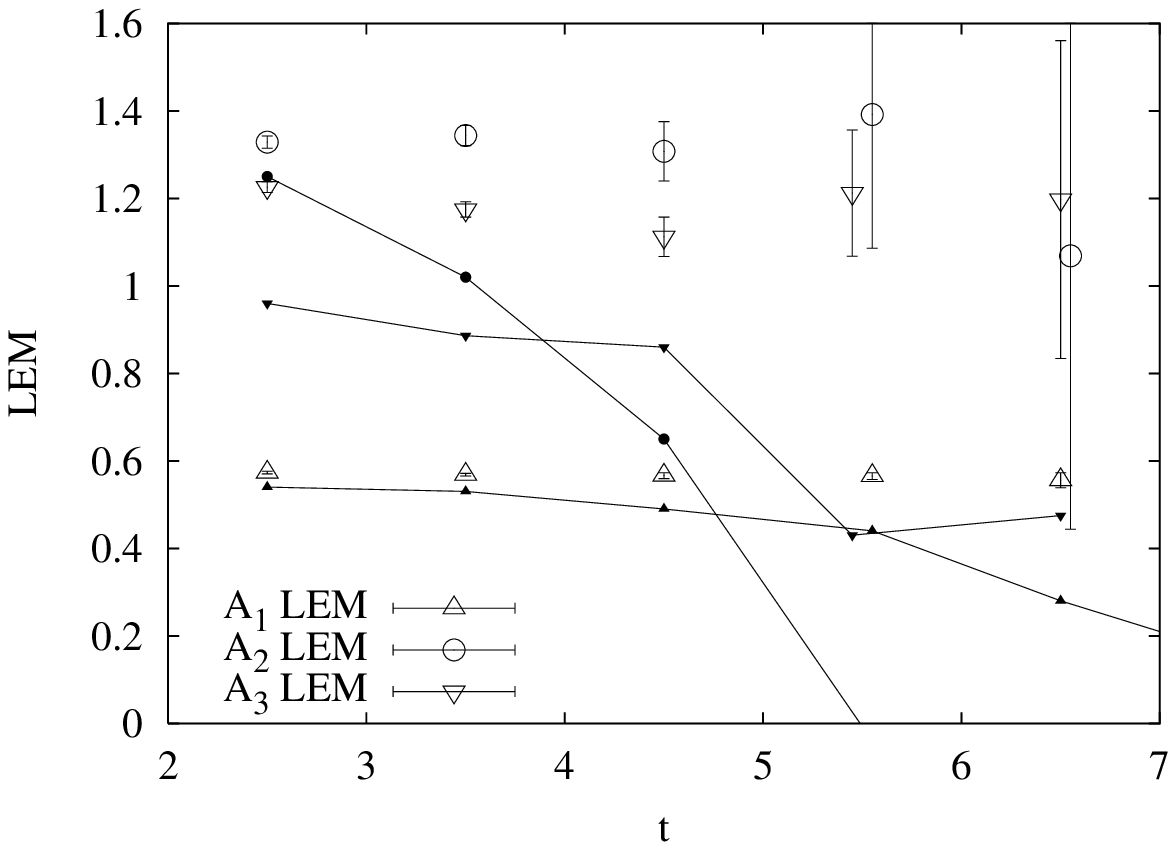,angle=0, height=9cm}	
    \end{minipage}}
\vspace*{0.5cm}

\caption[a]{The local-effective-mass of various correlators, and of the 
variance  on the latter, as function of the Euclidean-time separation $t$. 
The distance between fixed time-slices is $\Delta = 8$ and 
the number of measurements under fixed boudary
conditions is $n=1600$ for the top plot and $n=200$ for the bottom plot.
For 2+1D $SU(2)$, at $\beta=12,~V=32^3$.}

\la{fig:LEM}
\end{figure}

\begin{figure}[tb]
\vspace{-2cm}
\centerline{\begin{minipage}[c]{11cm}
	\psfig{file=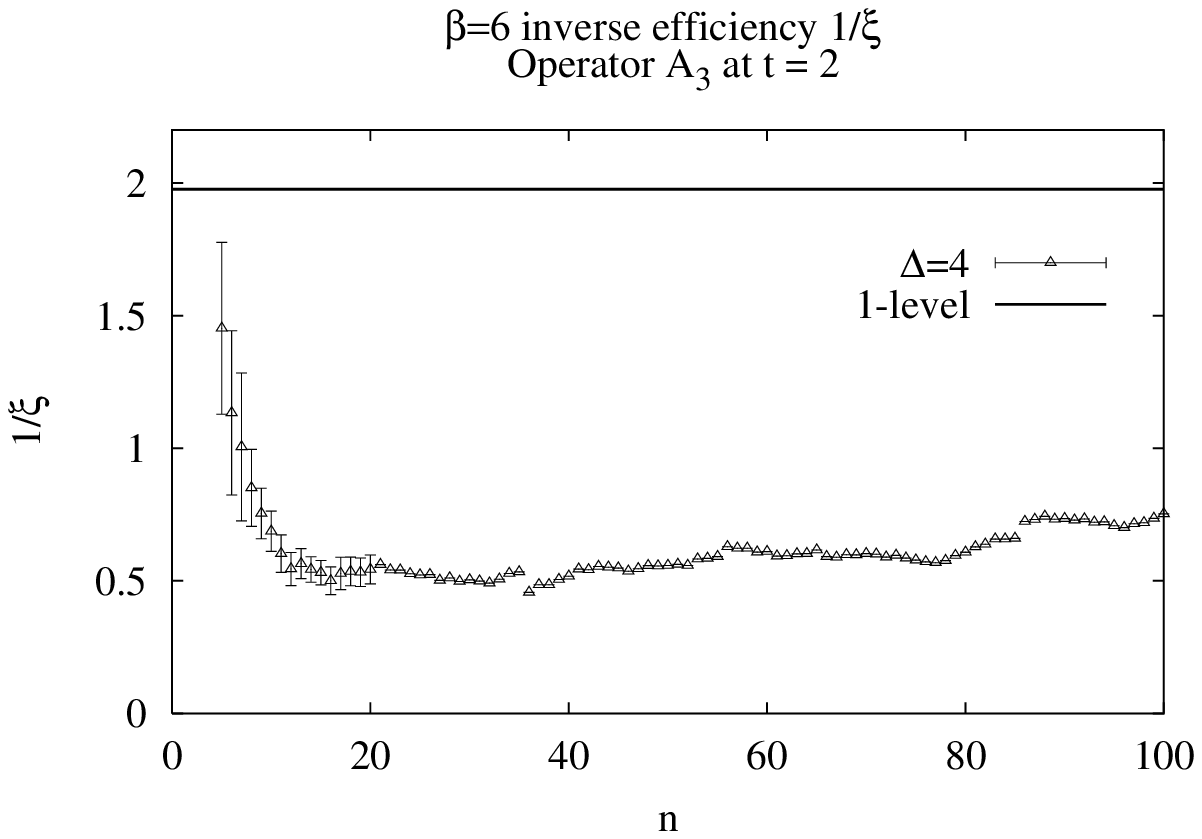,angle=0,width=10cm}\\
\psfig{file=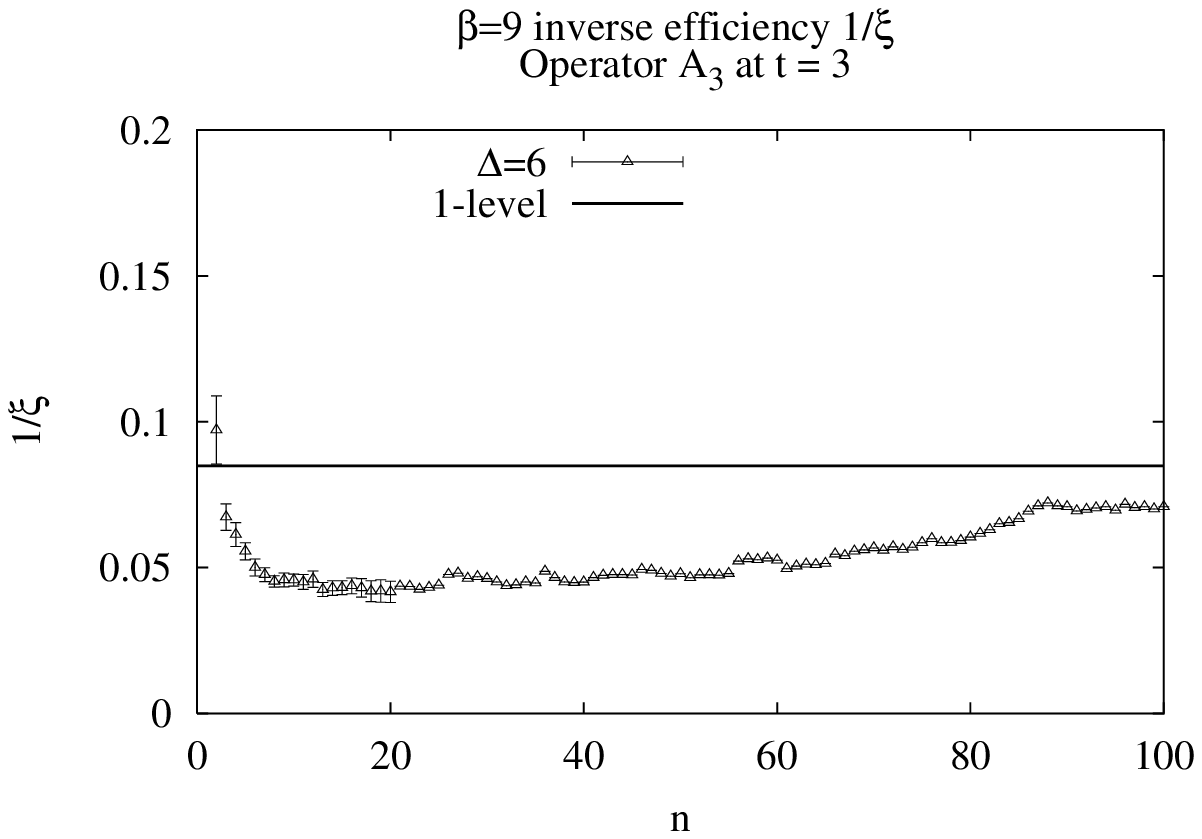,angle=0,width=10cm}\\
\psfig{file=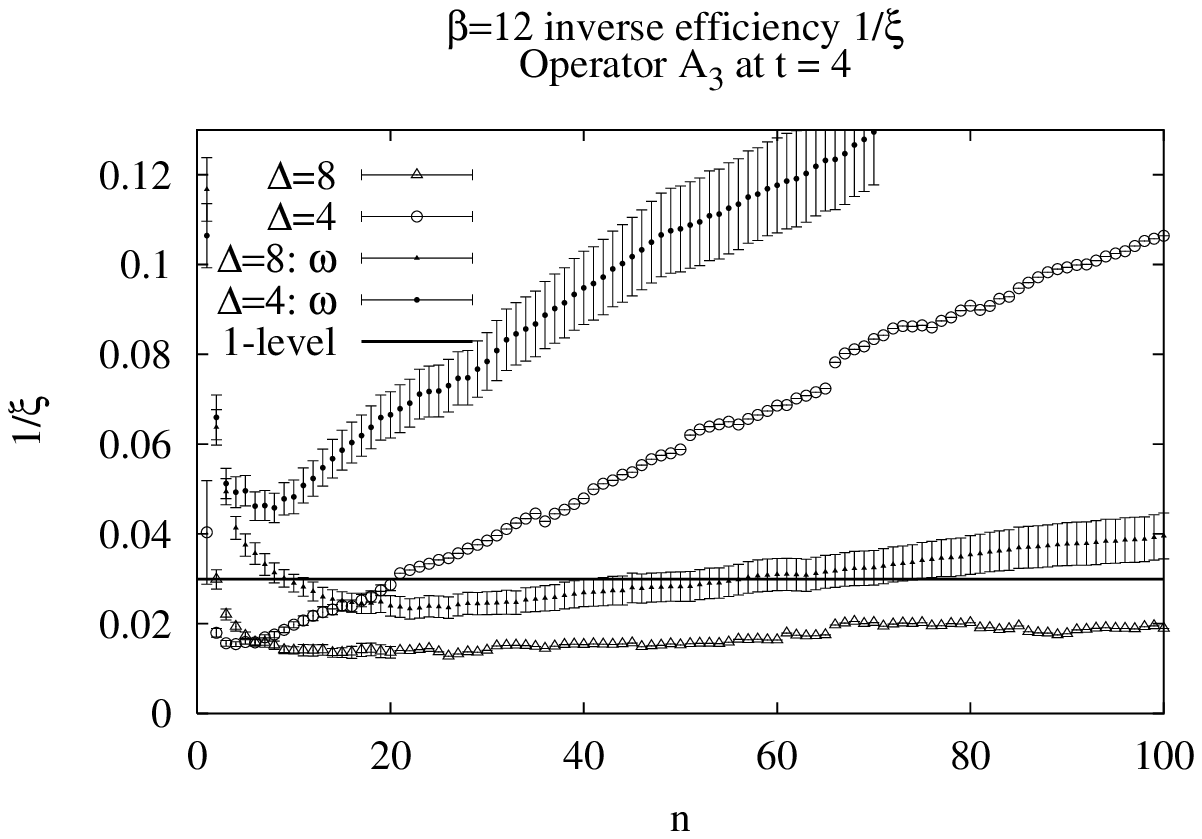,angle=0,width=10cm}
    \end{minipage}}
\vspace*{0.2cm}
\caption[a]{$A_3$ inverse efficiency and its predictor $\omega$ 
in 2+1D $SU(2)$.}
\la{fig:a3}
\end{figure}

\begin{figure}[tb]
\vspace{-2cm}
\centerline{\begin{minipage}[c]{11cm}
\psfig{file=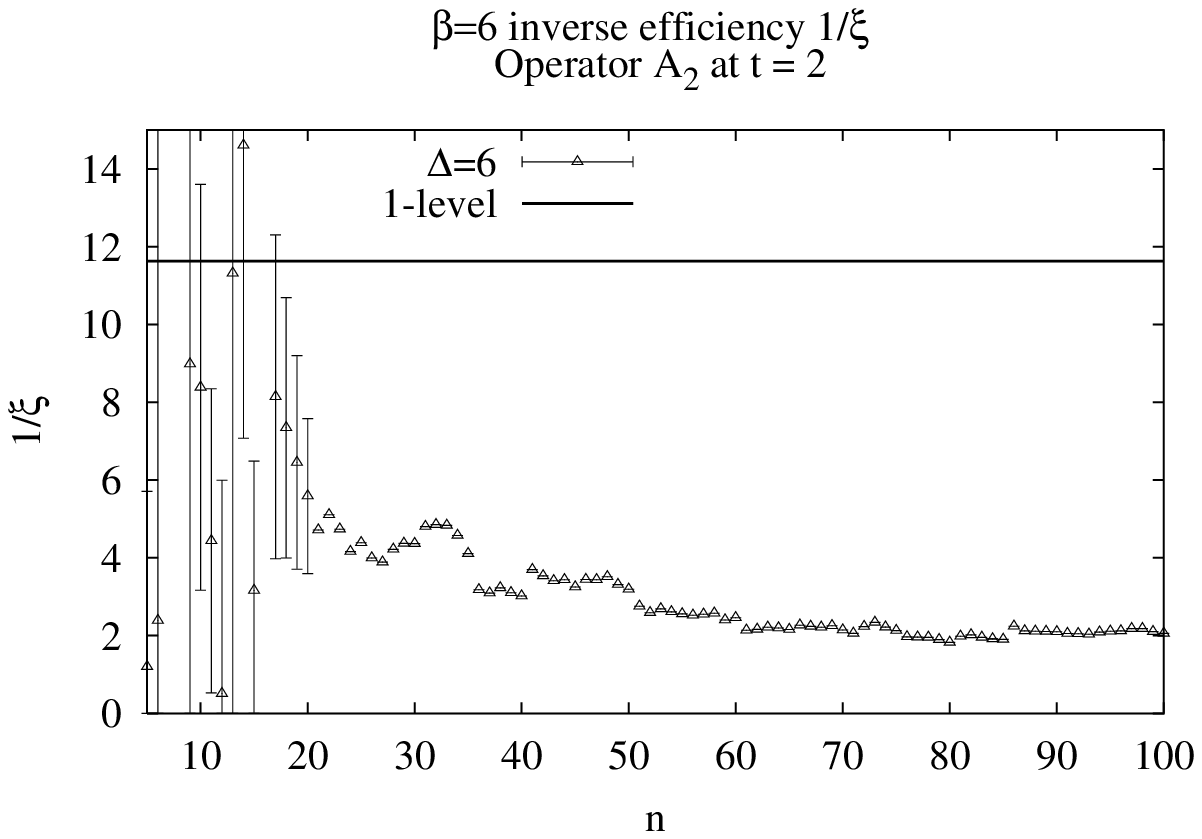,angle=0,width=10cm}\\
\psfig{file=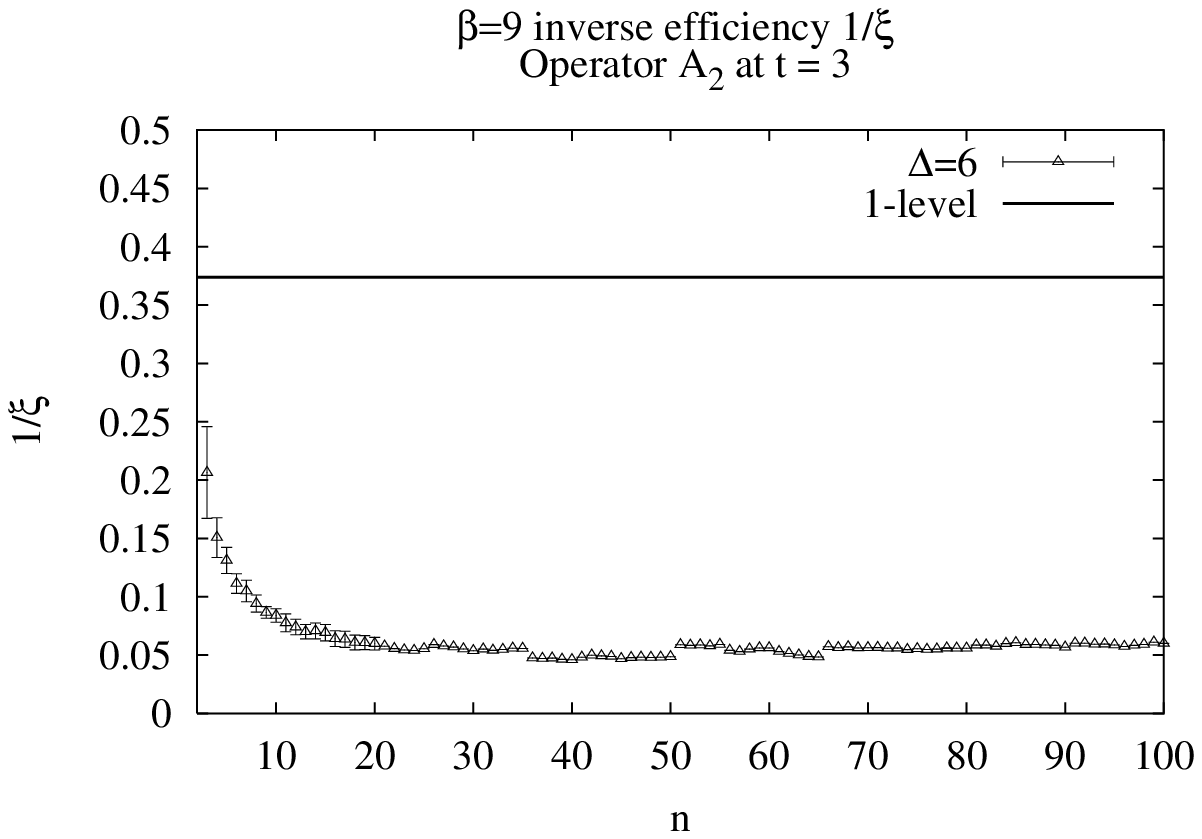,angle=0,width=10cm}\\
\psfig{file=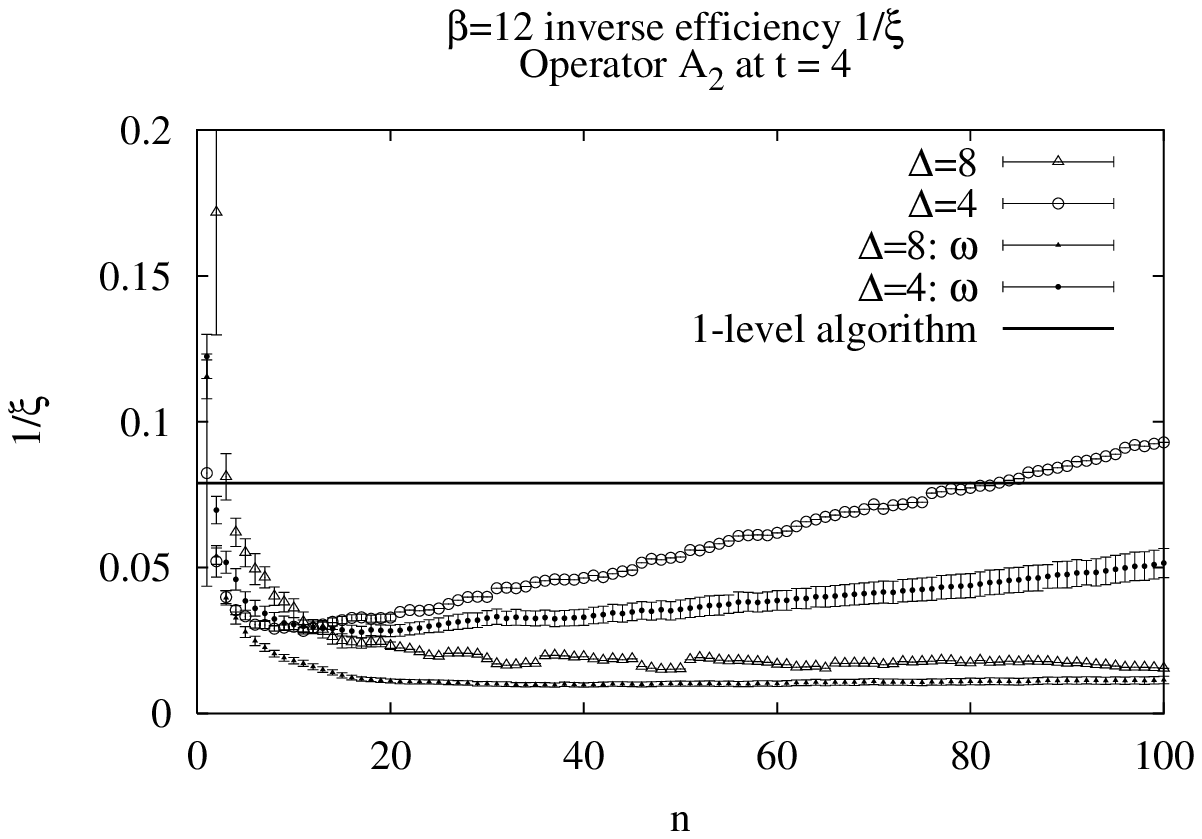,angle=0,width=10cm}
    \end{minipage}}
\vspace*{0.2cm}
\caption[a]{$A_2$ inverse efficiency and its predictor $\omega$ 
in 2+1D $SU(2)$.}
\la{fig:a2}
\end{figure}

\begin{figure}[tb]

\centerline{\begin{minipage}[c]{13cm}
        \psfig{file=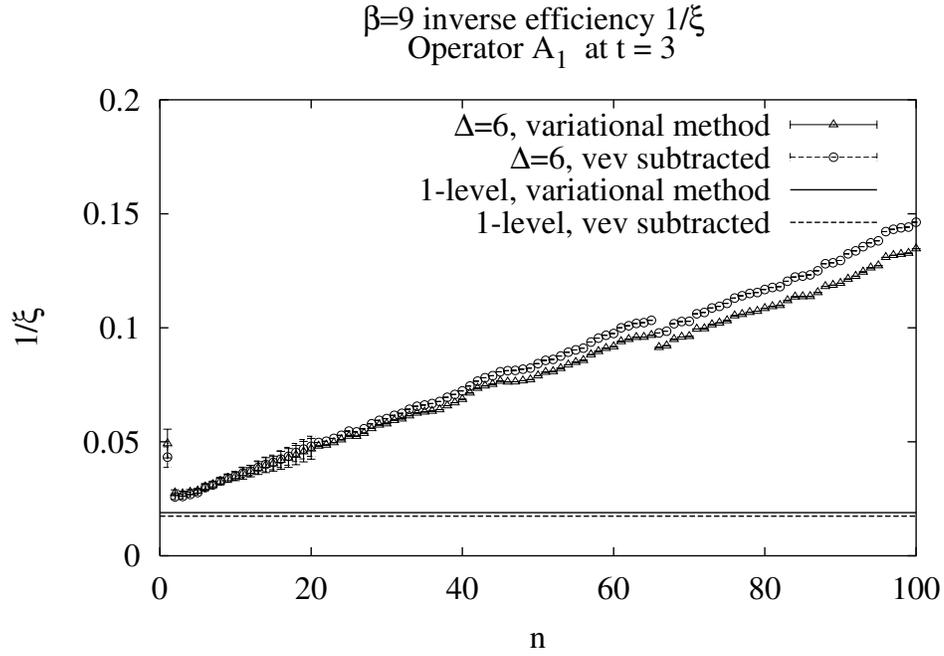,angle=0,width=13cm}\\
	\psfig{file=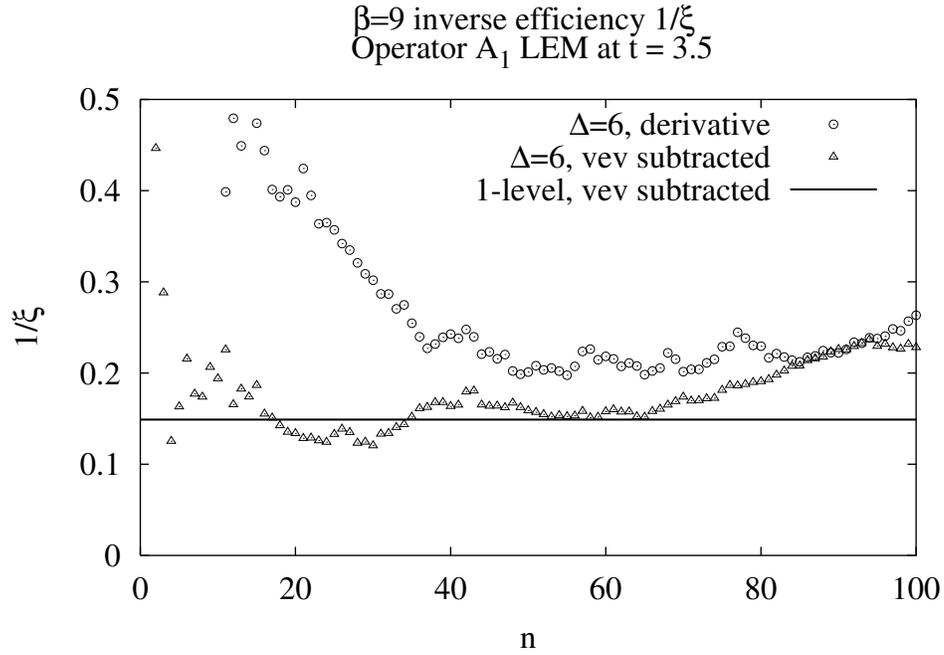,angle=0,width=13cm}
    \end{minipage}}
\vspace*{0.5cm}

\caption[a]{$A_1$ correlator (top) and LEM (bottom) efficiency curves  using
various methods of VEV subtraction. In 2+1D $SU(2)$.}

\la{fig:a1}
\end{figure}

\begin{figure}[tb]

\centerline{\begin{minipage}[c]{13cm}
       \psfig{file=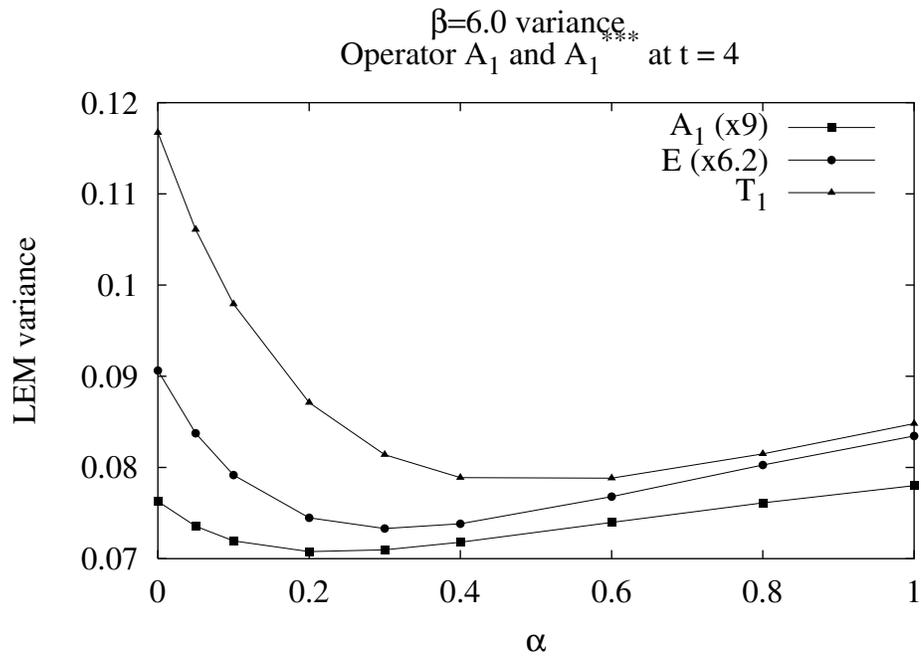,angle=0,width=13cm}
    \end{minipage}}
\vspace*{0.5cm}

\caption[a]{Local-effective-mass variance for three different states, as
a function of the weighting parameter $\alpha$ 
(cf. section~\ref{performance}). The $A_1$ and $E$ curves have been rescaled
as indicated. For 3+1D $SU(3)$ at $\beta=6.0$, $V=16^3\times36$.}

\la{fig:a1var}
\end{figure}
\end{document}